# Securing Music Sharing Platforms: A Blockchain-Based Approach

Adjei-Mensah Isaac[1]*, Isaac Osei Agyemang[1], Collins Sey[2], Linda Delali Fiasam[2], and Abdulhaq Adetunji Salako[3]

[1] *School of Information and Communication Engineering, University of Electronic Science and Technology of China*
[2] *School of Information and Software Engineering, University of Electronic Science and Technology of China*
[3] *School of Computer Science and Engineering, University of Electronic Science and Technology of China*
{iadjeimensah, agyemangisaac45, adetunjisalako}@gmail.com; clinsey2@yahoo.com; linda.delali92@outlook.com
*Correspondence:      iadjeimensah@gmail.com

*Abstract* – From online education and trading, all aspects of our lives are affected by digital technology. Among them, the storage of "music" has also entered the digital era. Music productions created by artists have brought great joy to people. However, when artists upload their works, they are most downloaded and reprinted by others, and copyright information and the issue associated with the sharing of music arise. This will have a significant negative impact on the enthusiasm and motivation of artists. This paper provides an internet database platform for artists, which uses the distributed and tamper-proof technology of the Ethereum blockchain to store music works, and protect the copyright information of each album or music produced by artists in the music industry. Design and implementation of the system model and data storage are proposed and data storage processes based on the Ethereum smart contract are demonstrated in detail. The system stores music information on the blockchain network, using the smart contract to provide artists with a fast and efficient royalty payment. Node.js is applied to carry out the experiments of our system, and we test Remote Procedure Calls (RPC) with available account and private keys for contract development and use block explorer to track music information on the blockchain. Our system enables copyright revenue to be attributed to music creators that will help to eliminate the illegal uploading of music on other websites.

*Keywords* – **Blockchain, Smart Contract, Online Music, Copyrights, File Sharing, Peer-to-Peer Networks.**

## 1. INTRODUCTION

Over the years there have been a significant number of downloads made in the music industry in diverse digital formats and extensions (e.g., mp4, mp3, etc.). However, the majority of the aforementioned downloads are acquired illegally yet indexed or accounted for in the global download digital library. Due to the illegal aspect of downloads, content creators lose much revenue and incur costs leading to the issue of copyright which this paper focuses on. In historical prevalence, it is worth noting that certain practices and the removal of copyrights in whatever digital form it presents itself will bring about a decrease in their price, broader distribution, removal of cost of transactions, and publishers being given less power (Ericsson, 2011).

Regardless of the conclusion in the preceding paragraph drawn by Ericsson, it is to say this article was ahead of time. The case of copyright is never an easy one. The purpose of this research is to pay tribute to Justice Breyer by exploring the effects of online music distribution. It is a maxim that seems to be a relative approximation of an environment less protected by copyright, to obtain valuable information on fundamental questions of copyright. For some time now, the widespread and evolving practice of online music distribution has changed the way we collect, consume (listen to), and experience music. This can be based on the fact that it cannot be done without the digital aspect of it. The comparison between new and old online music distribution has grown considerably over the years, as there appears to be a discrepancy between the copyright system and the constitutional goal of promoting the advancement of the art, scientific and practical aspects. The economic theory, therefore, played a role in legitimizing the American copyright system, into which economists were drawn by the debate over the scope of copyright protection. Economists have studied the creation and dissemination of works because they are linked to efficiency and social well-being.

In as much as this has been the case, economic research has also been directed into the amount of copyright protection needed to provide an economic incentive for the creator and dissemination of original work. The activities of the Recorded Music Industry (RMI) are strongly affected by Online Music Distribution (OMD). Technology in the advancement of digital content, storage, and distribution rendered the possibility of OMD on a large scale, RMI was still operating according to a relatively traditional business model. For many years, the innovative idea behind blockchain technology has led to a massive request for blockchain, mainly to remove third parties in the field of finance and to reduce the modeling based on trust (Nakamoto, 2008) and research. Since the advent of the technological market, blockchain has emerged as a promising technology. Blockchain was originally called the general ledger of the currency BTC (BitCoin) (Horváth et al., 2018)(Nakamoto, 2008). Blockchain is attributed to decentralization, immutability, traceability, and the creation and displacement of digital assets (Anjum et al., 2017). Blockchain eliminates the availability of intermediaries and/or central institutions. There is a single point of failure when the centralized servers are offline or inaccessible at a particular time or at all times (Anjum et al., 2017) (Dorri et al., 2016). In intermediaries and/or central institutions (Nakamoto, 2008), transactions must go through several nodes or exchanges to carry out final transactions with the designated node in the chain. Consequently, the advent of blockchain allows individuals/nodes to record information, and transactions are transported in the structure of a single unit without having to know the data that has to be exchanged in advance but to carry out transactions transparently and securely. If the internet in our various lives is currently being used widely, blockchain technology will stake a claim where it would be widely used in all aspects of life.

This research focuses on a file-sharing platform that tends to destabilize the traditional World Wide Web (WWW). The traditional WWW known globally as the internet holds several databases used for storing data and to some extent information. These databases are vulnerable to variant internet attacks. When such attacks have been achieved and global databases compromised, resources are lost and to some extent, a ransom is taken. In as much as the problems on the internet in the former statement is a part of our focus of the research, the main focus is music sharing by copyright holders. Taking a look across the internet, we have original works by content creators distributed across many databases without the consent of these original creators, hence we seek to create a platform where once a content creator uploads his/her work on the internet, it cannot be replicated or uploaded again unto the internet and hence the original copy will be made available to users of the internet.

## 2. RELATED WORKS

There are several ways to illegally download music and offenses: software piracy, copyright, theft, piracy, file-sharing (electronic and/or non-electronic methods), and free activation. Softcopying – copying digital files such as music or movies – is a type of software piracy in which a copy of legally licensed software violates the original license agreement. Unlike commercial hacking, softcopy copies the program to multiple users, rather than making copies for profit.

By establishing a research model, (Zeng, 2020) put together the digitization of music resources and the management of its copyright, and also thought that relevant authorities ought to enhance the credibility of digital music. In an era where digitization has become a goal to achieve, (Han et al., 2020) was of the view that copyrighted works of individuals and organizations cannot be fully protected, which will result in infringement made by people, but with the advent of blockchain which is a decentralized database the copyrighted works of individuals can be stored and protected.

In P2P file sharing, the cost or the price of a file is one of the most attractive factors in the implementation of piracy (Harcar, 2014), in Harcar's work the overpriced price of the invention and one of the main reasons for pirating customers. Despite the huge differences in quality between the original content (music) and the copied content, consumers do not see the deterioration in quality as a factor holding them back. Consumers have the same idea, so blaming music companies for overpricing music products is a major factor in participating in music piracy. In the P2P music sharing community, individuals do not see themselves as parties to counterfeit business interests, but rather as people who contribute to mutually beneficial social groups. Chowdury et. al (Michel, 2007) although Napster closed in 2001, opponents of file sharing claim that indiscriminate copying on the internet will reduce sales, while those who support it say file-sharing is harmless and may even increase music sales. Napster is often described as peer-to-peer software for music distribution. Napster users have downloaded their proprietary software from napster.com. With users actively online, Napster then created a floating party for music distribution, then they compiled a centralized song list from users for the availability of download from other users. Napster was not a music store. Napster is an intermediary in the main aspect, it is a personal matchmaker who wants to exchange songs. But between 1999 and 2003, file-sharing reduced sales of music albums by more than 13%, which is clear evidence that file-sharing did not lead to the widespread popularity of music purchases. In the works of (Guo et al., 2020), they identified some issues associated with copyright such as the difficulty in confirming rights of ownership, authorization, and maintenance rights in the traditional format. In this regard, they were of the view that the rate at which data is transmitted can be improved upon with the help of self-supervision, traceability, and decentralization of blockchain technology through a multi-channel medium to obtain an accuracy of time complexity. In providing proof of the existence of piracy, (Peng et al., 2019) in their work built on the InterPlanetary File System looked at the consistency of watermarks when information is being transacted (shared) on a chain.

A study has been conducted to find out if Spotify's music streaming service has succeeded or failed in removing consumers from illegal file-sharing. This is necessary because organizations and artists are aware of the Spotify service. Some countries have served as the basis for whether Spotify has reduced privacy in these areas. The research methods used are quantitative research and meta-analysis. The countries concerned suspect Spotify of cracking down on music piracy and encouraging the public to avoid illegal file-sharing through incentives. However, it can be observed that Spotify compensates for the loss of revenue caused by piracy, which in turn contributes to the stability of the music industry. Ping et. al (Lai et al., 2009) proposed a system to leverage the features provided by high-performance interconnections to improve the capabilities of File Transfer Protocol (FTP) in high-end data centers. Igmar and Sebastian (Baumgart & Mies, 2007) proposed a key-based routing protocol using Kademlia. This protocol uses parallel searches on several disjoint paths, uses password puzzles to limit the free generation of node Id, and introduces reliable homologous broadcasts, thus very resistant to common attacks. The latter requires that the data be stored in a secure place. Capastru et. al (Nformatics et al., 2017) proposed to use a combination of content analysis (CA) and web content analysis (WebCA) to allow WebCA to perform data entry. In their research, when the input data is based on online resources, their research methods cannot provide high reliability for scientific research, and the knowledge it is easy to confirm what follows from the method used.

## 3. SYSTEM ARCHITECTURE

The architecture of our platform uses the Inter-Planetary File System (IPFS) which is a decentralized file-sharing platform. In IPFS, involved parties have to establish a connection for the processes to be met. The feel of having close contact with a content creator brings with it a lot of desires that most fans will love to have. Every available decentralized application on the network will need to verify individuals who connect to the database for file retrieval. This serves as the basis on which individuals can get desired files. Music lovers' will need to be confirmed and authenticated by the usage of their login credentials. This will be through the usage of the email address and password used for registration. Collected data will be saved and encrypted on the IPFS platform. Access to the platform is done by verifying users' collected data from the blockchain. On a successful activity or request for a music file from the IPFS network, the activity is recorded to the blockchain. The recorded transaction is used to validate the revenue generated from the music file granted. Fig. 1 is a graphical representation of the proposed secured music sharing platform utilizing blockchain.

*3.1. Registration and Authentication*

The system begins with an execution having to do with the registration phase. A new block is added upon successful registration by a user. The data inputted such as the email address and password will then be cross-checked against the existing data. If it pre-supposes that the data is a new one, it is then hashed and added as a new block, after a consensus agreement has been met by all processes involved within the nodes. Once this consensus process is met, it is then broadcasted across the entire blockchain network, where these nodes uphold the availability of the blocks and attach them to the blockchain and thereby having the privilege and direct admittance into the blockchain network. This process is the same for authentication to take place. As a user's data is received for authentication, it is compared to the hash values of previous registration to check if it does not exist and upon the successful check, access will be granted. If user details do not correspond to the database, login is revoked and an attempt is made again.

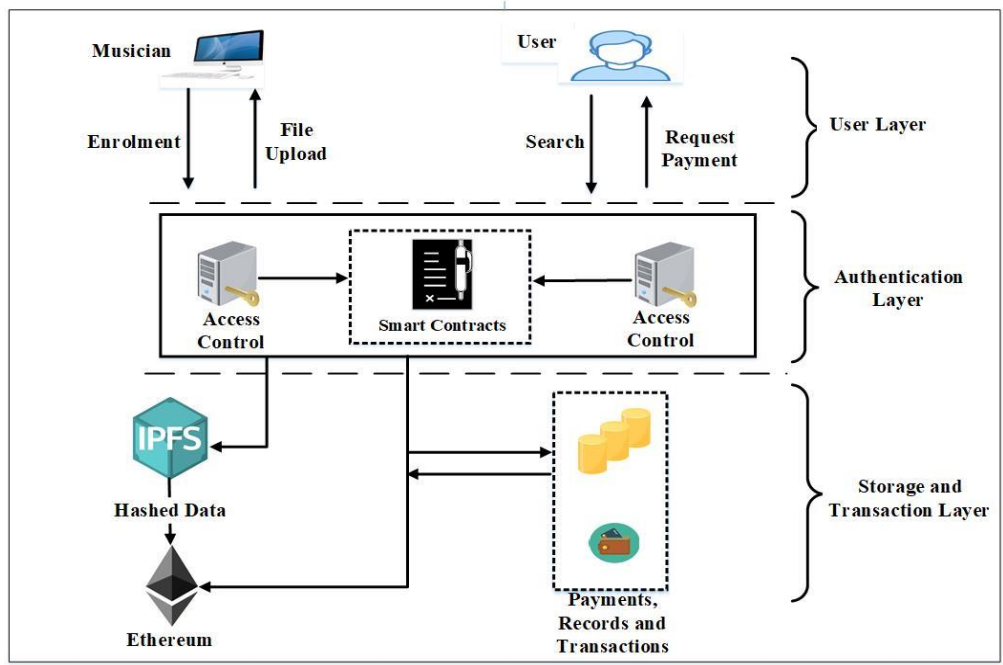

Fig. 1. Music sharing platform.

## 3.2 Consensus Mechanism

The consensus mechanism is to ensure that all nodes in the blockchain network are synchronized with each other and reach an agreement on legal transactions and added to the blockchain. To function properly, the consensus in the blockchain is considered crucial. The transaction is constantly verified and the blockchain is constantly reviewed by all the nodes. It should be mentioned that if there is no good consensus, the blockchain risks being subjected to various attacks. IPFS utilizes the principle of Proof of Stake (PoS). PoS uses an election process in which one node is randomly chosen to validate the next block hence no miners are needed to reduce higher computing power which is used by the Proof-of-Work (PoW) consensus mechanism, but validates where nodes do not mine the new block but mint them. Validators are not chosen completely in a random way, but by depositing an amount in the network as a stake. The size of the stake determines the chance of choosing a validator to forge the next block, which is linked linearly. This consensus is much better than the PoW where economics with higher computing power enjoy many rewards. Validators in a PoS check against all transactions within it to ascertain if they are valid before validating the next block and then sign off on the block if everything checks out by adding it to the blockchain.

## 3.3 Structure of a Parent Block

In designing a blockchain network, blocks involved are described as container-type data structures that tend to keep the transactions that need to be seen as an additional block to the blockchain. Blocks within a parent block of a blockchain are distinctively acknowledged by a file format that distinguishes each block from the other to achieve correspondence amongst themselves as illustrated in Fig. 2. The following arrangement representing the block size of all blocks in the network is the number of bytes arranged for the end of the block. Once the block size gets connected to the block header, the header which contains the metadata is used in verifying the authenticity of the block. The header is then identified by the hash generated by the SHA256 password hashing algorithm and guarantees invariance, which is a key security feature of blockchain technology. When changes are made to the block header, the hash also changes, resulting in changes to the hash of the former block and other corresponding blocks until a change is also made to the genesis block. Due to this complexity, it is difficult for an attacker to modify the content. Since the attacker cannot implement the change, it provides higher security.

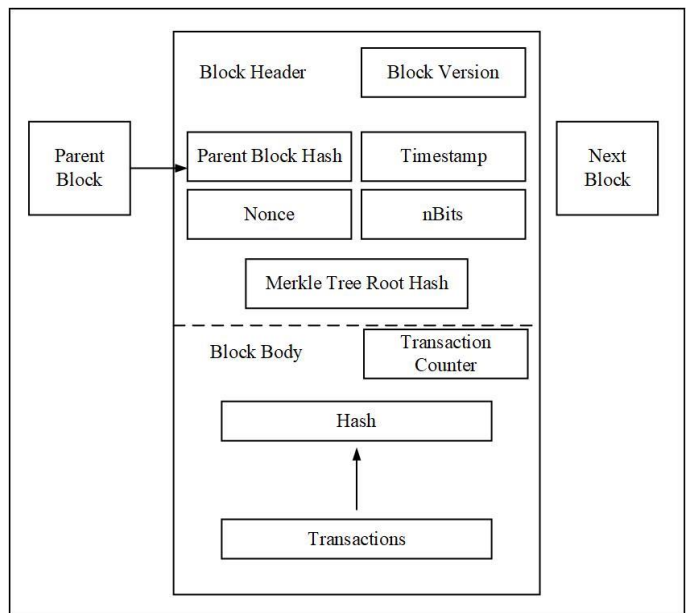

Fig. 2. Structure of a Parent Block.

The header contains the versioning of the block and provides confirmation rules which ensure compliance with the datatype. The version number is used as the support basis to indicate protocol decisions issued by minors, resulting in a specification of the type of data and its properties that are being processed. The hash (SHA256) of each previous block is confined to its block header. In the parent block of a blockchain, the Merkle root hash has its header described as a data structure to efficiently summarize all transactions in each block. It also ensures that no block in the blockchain network can be changed without affecting the header files. The Merkle tree is capable of generating digital fingerprints of all transaction sets. This is accomplished by continually hashing events in the blockchain and adding to the productivity of the current block. SHA256 is always applied twice, so it is called double-SHA256.

*3.4 Structure of a Side Block*

The side block of the block structure consists of the following format: part of the ID of the main block is added to the ID, which is created by the processing of IDs and consensus achieved on the side block. The next module is the block size, which is considered the entire block size (in bytes). Inside the side, the block is also a block header, which has six entities similar to the structure of the parent block, namely the version number, which is responsible for identifying the data type of the hash. Merkle root is used in the creation of the side block, the previous side block, and all the side blocks for the parent Events, timestamps, target difficulty, and random numbers. They are similar to the block of the parent structure but linked to the edge block. The sidebar consists of action counters, which record violations in a single report. In the box next to it is the transaction, which consists of Timestamp of Violation (TOV), User Identification (UID), Violation Type (VT), Node Identification (NID), and Node Signature (NS). The last of these components is chain time, which is in control of recording the time when a block arrives in the blockchain network and broadcasts across the network.

*3.5 IPFS, Ethereum, and Smart Contract Access Control*

In this section, descriptive information of IPFS is given, and also an analysis of how the Ethereum smart contract manages the permission and presents the necessary modifications made to IPFS that allow users to share music files is also illustrated. IPFS relies on a Distributed Hash Table (DHT) to find the nearest node to receive file location information. Ethereum allows the implementation of smart contract-based permissions or access with removal, addition, and changes in permission being recorded on the blockchain. The smart contract storing and managing the access is designed with Solidity programming language which has implicit functions like $msg.sender$ which contains the addresses of the user sending the transaction. The design of a smart contract access control is done with two generic functions: $MusicOwner$ and $chkAccess$.

## 3.6 Instances on the IPFS, Ethereum, and Smart Contract Access Control

**$addBlock$**: This function is used to register an IPFS music file chunk in the access control. Cryptographic hashes are used to identify chunks. The function accepts the computed hash as its argument. It checks to ensure that it's not empty and also the hash does not exist for another owner. The contract calls revert if either of these checks fails. Otherwise, the hash of the chunk is used as a key in the mapping ($fileMapping$), using the value as a new instance of the FileData structure, which has its value for owner set to $msg.sender$. This new data is committed to the contract's storage.

**$grantAccess$**: This takes an address and a byte32 variable representing a chunked hash as an argument. If the address is empty or the hash is empty, it reverts. Otherwise, it recalls the corresponding data from storage by accessing $fileMapping$, with the key being the hash argument. It checks to ensure that $msg.sender$ correspond to the owner recorded in the instance of FileData accessed through $fileMapping$. If it does not match, the call is reverted. If it matches, the access mapping has the address argument added as a key and the corresponding boolean is set to $True$. Finally, the address argument is added to the $allowedAddresses$ array and kept in the contract's storage.

**$removeAccess$**: This function operates in the same way as $grantAccess$ function, the only difference being that it sets the boolean in $fileMapping$ to $False$.

**$MusicOwner$**: Takes bytes of data representing a chunk hash argument, it returns $False$ if this is an empty value. It returns $True$ on the condition that $msg.sender$ equals the owner listed in $fileMapping$ with the chunk hash as the key, and $False$ if otherwise.

**$chkAccess$**: This takes one address and some bytes value representing a chunked hash as arguments. It returns true if the address argument matches the owner and false if either value is empty and/or if the address has previously been granted access. If the address has not been granted access, false is returned. The IPFS Access network is graphically illustrated in Fig. 3. The modified IPFS protocol interacts with the permissions package, which interacts with the smart contract through a node on the blockchain. It can send transactions to the smart contract that has been designed.

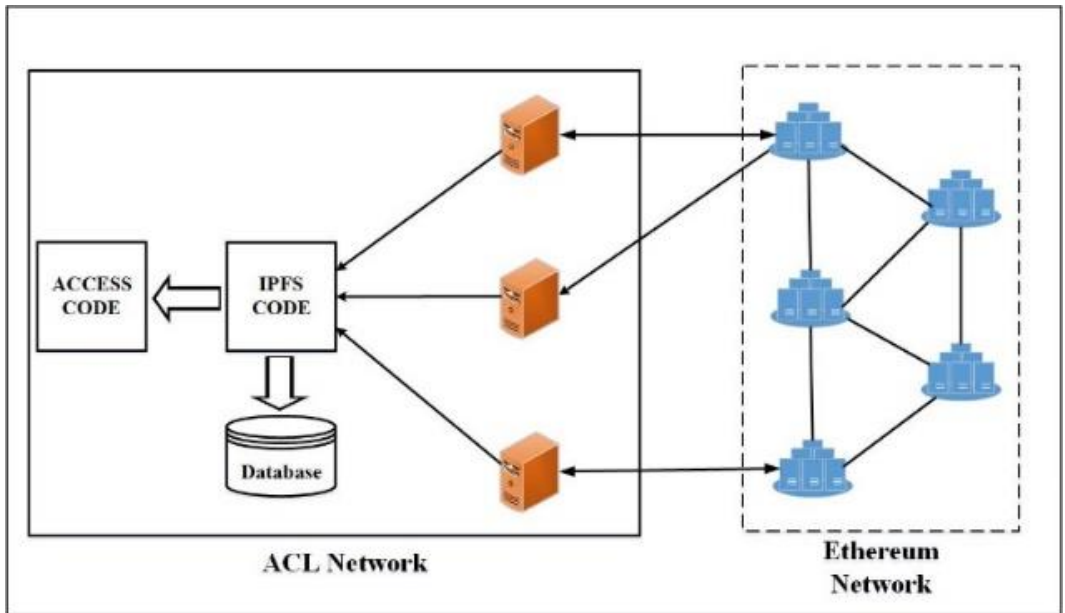
Fig. 3. IPFS, Ethereum, and Smart Contract Access Control.

***Uploading a music file***: Whenever a user uploads a file, the IPFS creates the chunks and the associated Merkle DAG. All descriptions associated with the music file are collected. Permission packages are passed, which checks for the already existing hash of the file uploaded. If no record is found, it adds the file in the smart contract, with the $addBlock$ function. This is done by forming the transactions necessary for this purpose.

***Downloading a file***: The IPFS Access Network allows copyright owners to easily grant access to specified users when the necessary payment has been made through the smart contract. A transaction is created and sent to the blockchain node once this condition is met. If the copyright owner grants the necessary permission, a node can request and get the chunks corresponding to the music file. IPFS then determines the holders of these chunks on the network and connects to them to get the chunks required to reassemble the file.

## 4. SIMULATION PLATFORM, DESIGN, AND IMPLEMENTATION

### 4.1 IPFS Daemon Start-Up

Content creators upload files and metadata through the web client. The tools for the web client ensure efficient functioning, Fig. 4 is an illustration of such an instance. The server-side is responsible for finding the port for file transfer by using HTTP responses and then saving the file at the localhost. The server-side then transfers the files to the IPFS and then retrieves the hash of the corresponding file. The metadata is used to bring out the content of the original file to be hashed into binary characters. After this, both the digital fingerprint and hashed characters are established on the server-side. The hashing key mechanism is achieved and the value is re-decoded after combining the jumble typescripts and metadata information. The generated key value of the original file is then uploaded onto the network. The data in the IPFS network is then synchronized and updated across the database of the nodes in the network. When other users of the system upload the same file and its metadata already exists in the system, copyright will set in thereby creating an error to the user that it already exists and there's a copyright violation. But on the other hand, if the metadata of the music file does not exist, it is then sent across

**Algorithm:** Pseudocode for Generating File Fingerprint String and Storing Music

**Input**: Music File
**Output**: fingerprint-string

1. **procedure** MUSICFINGERPRINT(music-file)
2.     FrequencyBlockSet{} = divide audio frequency into blocks;
3.     **foreach** block ∈ FrequencyBlockSet do
4.         subscript ← apply Fourier transform and extract subscripts
5.         computefingerprint(subscript)
6.         fingerprint-string ← compute fingerprint(subscript)
7.     **return** fingerprint-string
8. **end procedure**
9. **procedure** STOREMUSIC (fingerprint-string, music-file)
10.     **get** all server -fingerprint-strings()
11.     initialize copyright- exist {}
12.     **foreach** server–fingerprint ∈ server -fingerprint-string
13.         compare fingerprint with all sever finger-prints
14.         **if** fingerprint-string = server–fingerprint
15.             copyright-exist ← fingerprint-string
16.         **end if**
17.     **end for**
18.     **if** copyright- exist {} is empty
19.         store fingerprint-string and music-file
20.         send fingerprint-string to blockchain
21.     **End if**
22. **end procedure**

```
Administrator: Command Prompt - ipfs daemon
Repo version: 11
System version: amd64/windows
Golang version: go1.16.6
2021-09-12T00:06:15.543+0800    [31mERROR[0m  p2pnode libp2p/discovery.go:46  mdns error: No multicast listeners could
 be started
Swarm listening on /ip4/10.16.0.210/tcp/4001
Swarm listening on /ip4/10.16.0.210/udp/4001/quic
Swarm listening on /ip4/127.0.0.1/tcp/4001
Swarm listening on /ip4/127.0.0.1/udp/4001/quic
Swarm listening on /ip4/169.254.15.68/tcp/4001
Swarm listening on /ip4/169.254.15.68/udp/4001/quic
Swarm listening on /ip4/169.254.152.213/tcp/4001
Swarm listening on /ip4/169.254.152.213/udp/4001/quic
Swarm listening on /ip4/169.254.209.142/tcp/4001
Swarm listening on /ip4/169.254.209.142/udp/4001/quic
Swarm listening on /ip4/169.254.55.150/tcp/4001
Swarm listening on /ip4/169.254.55.150/udp/4001/quic
Swarm listening on /ip4/192.168.1.101/tcp/4001
Swarm listening on /ip4/192.168.1.101/udp/4001/quic
Swarm listening on /ip4/192.168.56.1/tcp/4001
Swarm listening on /ip4/192.168.56.1/udp/4001/quic
Swarm listening on /ip6/::1/tcp/4001
Swarm listening on /ip6/::1/udp/4001/quic
Swarm listening on /p2p-circuit
Swarm announcing /ip4/110.184.179.5/udp/27254/quic
Swarm announcing /ip4/127.0.0.1/tcp/4001
Swarm announcing /ip4/127.0.0.1/udp/4001/quic
Swarm announcing /ip4/192.168.1.101/tcp/4001
Swarm announcing /ip4/192.168.1.101/udp/4001/quic
Swarm announcing /ip6/::1/tcp/4001
Swarm announcing /ip6/::1/udp/4001/quic
API server listening on /ip4/127.0.0.1/tcp/5001
WebUI: http://127.0.0.1:5001/webui
Gateway (readonly) server listening on /ip4/127.0.0.1/tcp/8080
Daemon is ready
```

the entire network for an exhaustive examination to be done. If it does not exist, then the metadata is shared across the entire nodes on the blockchain network.

## 4.2 Current Files in Simulation Platform

Our database consists of the name of the content creator, title, date, meta hash, file hash, downloads made, and the file option which is used for downloading the music hash as shown in Fig. 5, it is timestamped for one to decipher between when a song was added. Files added by the content creator are hashed in the database, although the title of the music is available for any user to know the exact song being downloaded. Files already added cannot be added by another content creator as a result of the pre-existing metadata, and this brings about violations of copyright as a result of a check of hashes against the database to crosscheck if the metadata being added already exists, if it doesn't the metadata of the new song is then added, timestamped and hashed into the four nodes of databases in the blockchain network.

## 4.3 Download of Music File

When a music file is requested for download, the smart contract ensures that the requirement needed to be fulfilled is met before the file download pop-up is sent to the user for final download as shown in Fig. 6. If by any chance the user should lose the music file and needs to re-download the file, the user needs to meet the terms of the smart contract before the file download is initiated. From Fig. 7 (revenue mobilization), the revenue of each file (song) is calculated automatically and the number of downloads and the income generated are depicted. In this simulation, every music file is set to be downloaded for US$1.37.

| # | Author | Title | Date | Meta_Hash | File_Hash | Downloads | File Option |
|---|--------|-------|------|-----------|-----------|-----------|-------------|
| 1 | Johnson | AGAINST ALL ODDS-WESTLIFE | 26-Feb-2020 06:03:12am | 61e2f6e91acb2fb87e61da96a49ee23c | 2e9c08c8f3ee82ce6de907cbd61cc3e9 | 2 | Download |
| 2 | Sean Kingston | WHY YOU WANNA GO | 21-Feb-2020 17:15:11pm | cf70f4f87789a87da096ed921887c1ed | 81612184b048d0cc39c50ae907169c4d | 1 | Download |
| 3 | Wyclef | PERFECT GENTLEMAN | 21-Feb-2020 11:32:24am | 3eab9af2a99467943e96adf416c43433 | c5f639d5591b54dc4ba5ab8459b3b2f2 | 2 | Download |
| 4 | Stonebwoy | TUFF SEED | 21-Feb-2020 11:13:11am | be18fcfbe91edc311973ad61a545a93f | 25c9d05771ff5b18e06f0f036b5f9b61 | 22 | Download |
| 5 | Michael | STREET HUSTLE | 20-Feb-2020 17:51:56pm | 4c3fbb51228b35333ecf352ffeed2645 | c2a220e1dddd31bf6fd947fc14d90193 | 602 | Download |

Fig. 5. Uploaded music to the simulation platform.

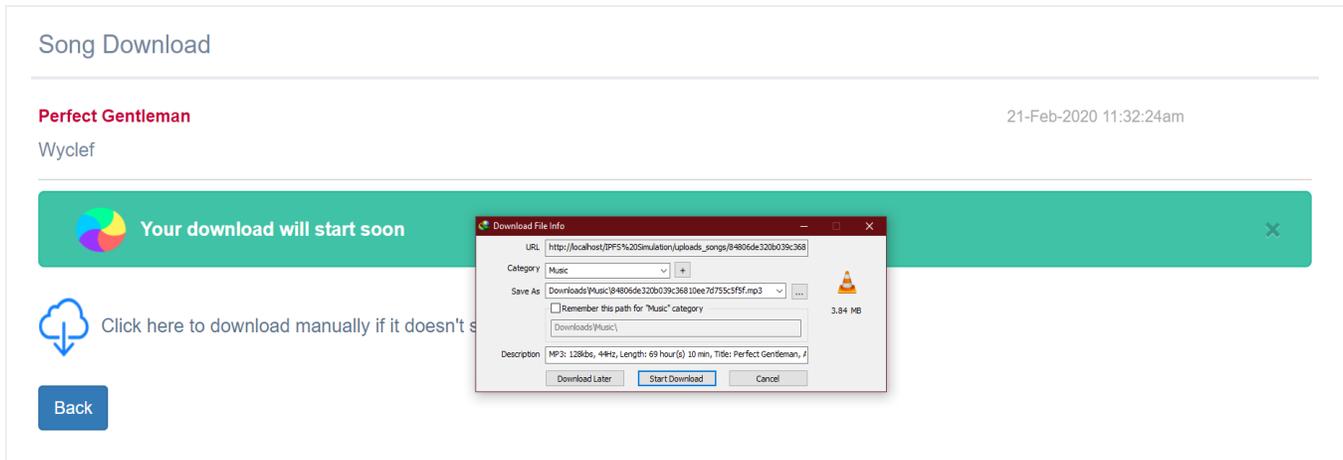

Fig. 6. A song in its download state.

## 5. SIMULATION RESULTS

In reality, cloud-based music copyright records services like IPFS not only enable the music industry to efficiently and flexibly manage music files with the aid of peer-to-peer storage services, but also make a great contribution to revenue generation and dispute resolution in the music industry. IPFS relies on a DHT to find the nearest node to receive file location information. Its ability to define how files move across the network makes it suitable for the proposed design. Since it's not feasible to store large amounts of data on the blockchain due to huge gas costs, IPFS serves as a great bridge to this gap. It makes use of cryptographic hashes which can easily be stored on the blockchain. IPFS protocol is leveraged to efficiently store and manage shared access to files on the network. A modified design of the IPFS protocol is presented that makes use of Ethereum smart contracts to provide access control to adding and downloading music files on the IPFS Network. The smart contract is used to maintain the access control to add files and also manage revenue from downloads on files. IPFS interacts with the smart contract whenever a file is uploaded, downloaded, or transferred.

### 5.1 Efficiency and Performance

After a thorough consideration of the proposed architecture, the ordinary question that comes into mind is, how will this eliminate copyright infringement in the music industry, and how revenue will be solely given to the content creator? In this section of the paper, a basic discussion of the idea on which this model is efficient is presented. The architecture presented earlier, refer to Fig. 2, demonstrated that the system is running on a public blockchain. As a result of the deployment on a public blockchain, data immutability is ensured. In ordinary layman's terms means the data will be available across the chain at all times even if there is an incident where a node does not exist (compromised). Using the decentralization attribute of the blockchain recordkeeping of music files will narrow the chances of original content being duplicated on the music sharing platform. Users being able to download music files does ensure that music files that smart contract has been reached and agreed upon and also music metadata should not be able to be added if it already exists as compared to other systems where duplicate files can be added.

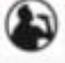

Fig. 7. Revenue mobilization.

## 6. CONCLUSION

This paper proposes a secured music-sharing platform that is based on blockchain and IPFS architecture. Our proposition eliminates and minimizes illegal music sharing of content creators across the internet using blockchain technology which also facilitates the checks of duplicates of metadata on the internet. This network of file sharing is constructed based on the Ethereum blockchain which uses a better way of consensus mechanism to achieve the stated goals of a smart contract in a fast and secured manner. Our simulation presents the various steps needed to visualize the operation of the proposed system. Our simulation introduced the registration and access control feature added to the IPFS protocol to ensure that music files are copyrighted. The smart contract ensures that the requirements needed to be met before accessing the music file are enforced. The tamper-proof state of the network was demonstrated such that every participating node ensured that records kept on downloaded music files correspond to the expected revenue. Effective accountability on revenue is achieved with our proposed system.